\documentclass[a4paper,11pt]{article}
\usepackage{jheppub} 
\usepackage{lineno}

\usepackage{epsfig}
\usepackage{amssymb}
\usepackage{amsmath}
\usepackage{lipsum}
\usepackage{graphicx}
\usepackage[dvipsnames]{xcolor}
\usepackage{tikz}
\usepackage{bm}
\usepackage{subfigure}

\usepackage{soul}
\usepackage[version=4]{mhchem}
\usepackage{multirow}
\usepackage{makecell}
\usepackage{array}
\makeatletter
\def\@fpheader{~}
\makeatother

\title{\boldmath Azimuthal decorrelation in diffractive dijet production }

\author[a,b,c]{Ding Yu Shao,}
\author[d,e]{ Yu Shi,} 
\author[f]{ Cheng Zhang,} 
\author[d]{ Jian Zhou,} 
\author[d]{ Ya-jin Zhou}

\affiliation[a]{Department of Physics, Center for Field Theory and Particle Physics, Key Laboratory of
Nuclear Physics and Ion-beam Application (MOE), Fudan University, Shanghai, 200433, China}
\affiliation[b]{Key Laboratory of Nuclear Physics and Ion-beam Application (MOE), Fudan}
\affiliation[c]{Center for High Energy Physics, Peking University, Beijing 100871, China}
\affiliation[d]{Key Laboratory of Particle Physics and Particle Irradiation (MOE), Institute of frontier and interdisciplinary science, Shandong University, Qingdao, Shandong 266237, China}
\affiliation[e]{CPHT, CNRS, \'Ecole polytechnique,  Institut Polytechnique de Paris, 91120 Palaiseau, France}
\affiliation[f]{School of Physics, Hangzhou Normal University, Hangzhou, Zhejiang 311121, China}

\emailAdd{dyshao@fudan.edu.cn}
\emailAdd{yu.shi@polytechnique.edu}
\emailAdd{chengzhang@hznu.edu.cn}
\emailAdd{jzhou@sdu.edu.cn}
\emailAdd{zhouyj@sdu.edu.cn}

\abstract{
We calculate the azimuthal angular decorrelation of diffractive dijets in ultra-peripheral heavy-ion, $ep$, and $eA$ collisions to probe non-perturbative diffractive transverse momentum-dependent distributions. Focusing on the dominant semi-inclusive channel with an unobserved semi-hard gluon, we perform an all-order resummation of soft gluon emissions for the transverse energy-energy correlator observable, accounting for both initial and final state radiation. We also analyze heavy-quark pair production and demonstrate the sensitivity of the decorrelation to the jet axis definition. Finally, we provide numerical predictions for relevant kinematics at LHC UPCs, HERA, and the future EIC. Our results demonstrate that the acoplanarity of diffractive dijet production could serve as a promising probe of diffractive transverse momentum-dependent distributions.
}

\begin{document}
\maketitle
\flushbottom

\section{Introduction}\label{sec:intro}

Diffractive dijet production in lepton-nucleon and lepton-nucleus scattering serves as a useful probe of the internal structure of hadrons. In the kinematic limit of two hard, nearly back-to-back jets, this process provides a unique window into the multi-dimensional partonic landscape of the target. At moderate Bjorken-$x$, exclusive dijet production in electron-proton ($ep$) collisions offers sensitivity to gluon generalized parton distributions (GPDs)~\cite{Ji:2003ak, Belitsky:2003nz, Braun:2002en, Braun:2005rg}. On the other hand, in the small-$x$ regime accessible at high-energy colliders, the focus shifts to gluon saturation~\cite{Gribov:1983ivg, Mueller:1985wy, Mueller:1989st, McLerran:1993ni, McLerran:1993ka, McLerran:1994vd, Iancu:2003xm, Gelis:2010nm}. Furthermore, diffractive processes also provide direct access to the gluon Wigner distribution, which encodes the full phase-space information of partons, including their position and momentum correlations~\cite{Hatta:2016dxp, Altinoluk:2015dpi, Zhou:2016rnt, Hagiwara:2017fye, Mantysaari:2019csc, Mantysaari:2019hkq, Boer:2021upt, Bhattacharya:2026qnd, Bhattacharya:2023hbq}.

Theoretical treatments must rigorously distinguish between exclusive production and semi-inclusive channels. Recent studies show that the cross section is dominated by coherent semi-inclusive processes~\cite{Iancu:2021rup, Iancu:2022lcw, Iancu:2023lel, Hauksson:2024bvv, Iancu:2025jsu} in the large di-jet invariant mass region. In this scenario, often termed diffractive tri-jet production, the primary production mechanism involves the diffractive production of two hard jets accompanied by a semi-hard gluon emission, typically collinear to the beam direction. This channel avoids the color transparency suppression characteristic of the exclusive process, making it the primary signal in experiments at the LHC, HERA, and the future EIC~\cite{ZEUS:2015sns, Accardi:2012qut, CMS:2022lbi}. The quark-antiquark pair in this tri-jet configuration remains in a color-octet state at short distances. Consequently, the system effectively scatters off the nucleus as a gluon-gluon dipole, resulting in a cross section that can be much larger than that of the exclusive channel~\cite{Iancu:2021rup, Iancu:2022lcw, Iancu:2023lel}.

An important observable for exploring the nucleon structure is the azimuthal correlation between the dijets. Conventionally, this is probed via the transverse momentum imbalance $\bm{q}_\perp$ and the angle $\phi_{qP}$, defined as the azimuthal angle between $\bm{q}_\perp$ and the relative jet transverse momentum $\bm{P}_\perp$.  Non-trivial azimuthal modulations in $\phi_{qP}$, such as $\cos(2\phi_{qP})$ asymmetries, act as unique signatures of the elliptic gluon Wigner distribution~\cite{Hatta:2016dxp, Altinoluk:2015dpi, Zhou:2016rnt, Boussarie:2018zwg, Hatta:2017cte, Mantysaari:2019csc, Mantysaari:2020lhf, Dumitru:2021mab, Hagiwara:2021xkf}. While previous analyses established that final state radiation (FSR) can mimic these asymmetries~\cite{Hatta:2020bgy, Hatta:2021jcd}, initial state radiation (ISR), which is kinematically allowed in the dominant semi-inclusive channel, tends to wash out these asymmetries~\cite{Shao:2024nor}. 

However, $\bm{q}_\perp$ is suboptimal for precision tomography due to experimental constraints. Its determination relies on jet energy measurements, which suffer from poor jet energy resolution and significant instrumental smearing. In contrast, the dijet acoplanarity $\phi_\perp = \pi - (\phi_1 - \phi_2)$ depends exclusively on azimuthal angles, benefiting from the superior angular resolution of modern detectors.  This makes $\phi_\perp$ a far more robust probe of the gluon intrinsic transverse momentum distribution inside a pomeron. 

To implement this acoplanarity measurement cleanly, the transverse energy-energy correlator (TEEC) provides an ideal framework~\cite{Basham:1978bw, Gao:2019ojf}. Unlike conventional jet observables that depend on clustering algorithms and axis determination, TEEC is defined via energy-weighted correlation functions of final-state particles. TEEC has been successfully applied to standard DIS and proton-proton collisions~\cite{Ali:1984yp, Ali:2012rn, Gao:2019ojf, Li:2020bub} and used to study saturation effects~\cite{Kang:2023oqj, Kang:2025vjk, Ganguli:2025aqa}. Yet, its application to diffractive processes remains largely uncharted. By formulating the TEEC for diffractive dijets, we provide a theoretically clean observable that reduces clustering ambiguities and enables a precise, factorization-motivated separation of the initial-state saturation signal from the Sudakov-induced radiative broadening.

From a phenomenological perspective, azimuthal decorrelations in photon-induced dijet and dihadron production have been widely studied within the small-$x$ framework as a probe of saturation effects over past decades (see, for example~\cite{Zheng:2014vka, Salazar:2019ncp, Kolbe:2020tlq, Taels:2022tza, Bergabo:2021woe, Bergabo:2022tcu, Iancu:2022gpw, Fucilla:2022wcg, Marquet:2025jdr}). In particular, recent studies in Ref.~\cite{Marquet:2025jdr, Gao:2026azd} show that heavy-mass effects suppress soft gluon radiation, leading to enhanced sensitivity of heavy-meson pair production to saturation effects and a clear mass hierarchy: the small-$x$ effect is stronger for $B$-meson pairs than for $D$-meson pairs.

In this work, we investigate the impact of soft gluon radiation on the acoplanarity observable in diffractive two-particle production. Adopting the diffractive transverse momentum dependent (DTMD) factorization framework~\cite{Iancu:2021rup, Iancu:2022lcw, Iancu:2023lel, Hauksson:2024bvv, Hatta:2022lzj, Hatta:2024vzv, Caucal:2023fsf, Rodriguez-Aguilar:2024efj} (also see Ref.~\cite{Lee:2025fml}), we perform an all-order resummation of soft gluon contributions. A central feature of our analysis is the consistent inclusion of ISR, which arises from the color-octet nature of the quark-antiquark pair at the Born level in the semi-inclusive process. By extending this insight to the TEEC observable, we show that ISR induces a significant broadening of the azimuthal correlation, which must be disentangled from the saturation signal. Furthermore, we expand the scope of our analysis to include heavy quark pair production. We also investigate the sensitivity of the observable to different jet axis definitions, contrasting the standard jet axis (SJA) with the winner-take-all (WTA) scheme~\cite{Bertolini:2013iqa, Neill:2016vbi}. 

This competitive interplay between ISR and FSR is critical for interpreting experimental data. Although diffractive processes in hadronic collisions are susceptible to factorization breaking effects, investigating this observable in UPCs provides a valuable way to probe the saturation regime at the highest available energies. Furthermore, the future Electron-Ion Collider (EIC) is essential for providing a rigorous test of the formalism in a clean environment. We present numerical predictions for UPCs at the LHC, while placing special emphasis on the prospective impact of the EIC. Our results demonstrate that the significant enhancement of acoplanarity driven by both ISR and FSR acts as a major background. This underscores the necessity of a comprehensive perturbative baseline to unambiguously isolate saturation signals in future studies.

The paper is organized as follows. In section \ref{sec:setup}, we outline the theoretical framework, reviewing the DTMD calculation of diffractive dijet production and presenting the factorization and resummation formalism. Numerical results are presented in section \ref{sec:num}. We conclude with a summary in section \ref{sec:conclusion}.

\section{Theoretical setup}\label{sec:setup}

We focus on the dominant semi-inclusive diffractive process, characterized by an asymmetric tri-jet configuration with the hierarchy $l_\perp \sim q_\perp \ll k_{1\perp} \approx k_{2\perp} \equiv P_\perp$. Here, the two-dimensional vectors $\bm{k}_{1\perp}$ and $\bm{k}_{2\perp}$ denote the transverse momenta of the two primary hard jets (with $P_\perp$ defining their characteristic hard scale), $\bm{q}_\perp$ represents the dijet transverse momentum imbalance, and $\bm{l}_\perp$ denotes the transverse momentum of the third, semi-hard gluon jet. In contrast to the purely exclusive channel, which is heavily suppressed by color transparency due to the color-singlet nature of the $q\bar{q}$ pair, this semi-inclusive process is distinguished by the emission of an additional semi-hard gluon. This emission leaves the $q\bar{q}$ pair in a color-octet state at short distances. Consequently, the short-distance system effectively scatters as a gluon-gluon dipole, circumventing color-transparency suppression and thus dominating the diffractive cross section in UPCs, HERA, and the future EIC~\cite{Iancu:2021rup, Iancu:2022lcw, Iancu:2023lel}.

Specifically, we analyze the coherent diffractive production of a dijet system in lepton-nucleus ($eA$) collisions and ultra-peripheral heavy-ion ($AA$) collisions. The underlying physical reaction is initiated by the fluctuation of a quasi-real photon into a quark-antiquark pair ($q\bar{q}$), followed by the radiation of a gluon ($g$) from either the quark or the antiquark. The resulting three-parton system ($q\bar{q}g$) then scatters coherently off the nuclear target via the exchange of a color-singlet pomeron. The corresponding kinematic process is denoted as
\begin{eqnarray}
\gamma(x_\gamma p)+A \rightarrow q(k_1)+\bar q(k_2)+g(l)+A,
\end{eqnarray}
where $\gamma$ represents the quasi-real photon emitted by the incoming lepton or a projectile nucleus with momentum $p$, and $A$ denotes the nuclear target. For $ep$ collisions, $x_\gamma$ is the longitudinal momentum fraction of the electron transferred to the photon. In UPCs, the source of the quasi-real photon can be either of the colliding nuclei. While the interference between photon emissions from the projectile and target, referred to as the double-slit interference effect, constitutes an intriguing quantum phenomenon~\cite{Klein:1999gv, Zha:2018jin, Xing:2020hwh, Mantysaari:2023prg}, its physical impact is strictly confined to the extremely low transverse momentum region ($k_\perp \lesssim 1/R_A \sim 30~\text{MeV}$). Because our analysis is centered on the semi-hard regime, where the pair transverse momentum is significantly larger, this interference effect can be safely neglected.

To parameterize the kinematics of this scattering process, we employ light-cone coordinates where the incoming photon propagates along the $+z$ direction with momentum $P^+$. The longitudinal momentum fractions of the photon carried by the final-state partons are given by
\begin{align}
    z_1=\frac{k_1^+}{x_\gamma P^+}, \quad z_2=\frac{k_2^+}{x_\gamma P^+}, \quad \text{and} \quad z_3=\frac{l^+}{x_\gamma P^+},
\end{align}
which strictly satisfy the longitudinal momentum conservation condition $z_1+z_2+z_3=1$. The corresponding momentum fractions with respect to the target nucleus (propagating in the $-z$ direction) are defined as $x_i = k_i^- / \bar{P}^-$. The exchanged pomeron carries a longitudinal momentum fraction $x_{\mathbb{P}}=x_1+x_2+x_3$. Furthermore, the condition for coherent diffraction requires $x_{\mathbb{P}} \ll 1/(2 m_N R_A)$, ensuring that the target nucleus remains entirely intact during the collision.

Our primary objective is to evaluate the distribution of the transverse momentum imbalance $\bm{q}_\perp=(q_x,q_y)$ of the two hard jets and their resultant azimuthal correlation. At the Born level, exact transverse momentum conservation dictates $\bm{q}_\perp = -\bm{l}_\perp$. Beyond leading order, however, soft gluon radiation substantially modifies this strict back-to-back kinematic configuration. The recoil generated by these soft emissions induces an azimuthal asymmetry, which is conventionally parameterized by the two-particle acoplanarity. We define the azimuthal angular difference as $\phi_\perp = \pi - (\phi_1 - \phi_2)$, where $\phi_1$ and $\phi_2$ are the azimuthal angles of the two primary jets, yielding an acoplanarity of $\alpha = |\phi_\perp|/\pi$. In the small angle limit, aligning the relative jet momentum $\bm{P}_\perp$ with the $y$-axis provides the geometric relation $\phi_\perp \approx q_x / P_\perp$. This relation establishes a critical theoretical link, allowing us to map the robust, experimentally clean acoplanarity measurement directly to the underlying $\bm{q}_\perp$ distribution governed by saturation dynamics and soft-gluon resummation.

\subsection{Factorization and resummation formula}

We now turn to the formalization of the acoplanarity observable. In the small-angle limit, the acoplanarity is straightforwardly extracted from the ratio of $q_x$—the component of the transverse momentum imbalance $\bm{q}_\perp$ perpendicular to the jet axis—to the hard momentum scale $P_\perp$. Within the framework of DTMD factorization, the corresponding $q_x$-dependent differential cross section is given by,
\begin{align}\label{eq:DTMD_fac}
  &\frac{d\sigma}{dq_x d^2 P_{\perp} dy_1 dy_2 }
  =\notag \\
  &\quad \quad \sigma_0  x_\gamma f_\gamma(x_\gamma) \int 
  \frac{d b_x}{2\pi} e^{i b_x  q_x} e^{-  \mathrm{Sud_a}(b_x)} \int d q_x' dq_y' \ 
  e^{-i b_x  q_x'}  \int \frac{d x_{\mathbb{P}}}{x_{\mathbb{P}}} x_g G_{\mathbb{P}}(x_g,x_{\mathbb{P}},q_\perp'). 
\end{align}
Here, $y_1$ and $y_2$ denote the rapidities of the two primary jets, $\sigma_0$ represents the Born-level partonic cross section, and $f_\gamma(x_\gamma)$ is the photon flux. Furthermore, $\mathrm{Sud_a}(b_x)$ denotes the Sudakov factor encoding the soft-gluon resummation, and $G_{\mathbb{P}}(x_g,x_{\mathbb{P}},q_\perp')$ is the DTMD, where $x_g$ is the momentum fraction of the pomeron carried by the active gluon.

Crucially, the resummation formula in this expression requires only a one-dimensional Fourier transform with respect to $b_x$, rather than the standard two-dimensional transform over the transverse coordinate space $\bm{b}_\perp=(b_x,b_y)$. This dimensional reduction is a direct physical consequence of the acoplanarity observable $\phi_\perp$ being exclusively sensitive to the orthogonal projection of $\bm{q}_\perp$ (defined here along the $x$-axis). By integrating out the unobserved parallel momentum component $q_y$, the conjugate Fourier variable $b_y$ is rigorously forced to zero via the identity $\int dq_y e^{i b_y q_y} = 2\pi \delta(b_y)$. Consequently, the soft-gluon resummation is dictated entirely by the dynamics along the $b_x$ direction, naturally establishing it as the exact conjugate variable to the azimuthal decorrelation.

To systematically unpack the components of the factorization theorem in Eq.~\eqref{eq:DTMD_fac}, we first specify the hard partonic scattering cross section for $\gamma+g \rightarrow q+\bar q$, evaluated as
\begin{eqnarray}
  \sigma_0=\sum_f \alpha_{e} \,\alpha_s \, e_f^2 \,  z_1 (1-z_1) \left [ z_1^2+(1-z_1)^2  \right 
 ]\frac{1}{P_\perp^4},
\end{eqnarray}
where $\alpha_e$ and $\alpha_s$ are the electromagnetic and strong coupling constants, respectively. The variable $e_f$ denotes the fractional charge of the active quark flavor $f$, and $z_1=(k_1^+)/(x_\gamma P^+)$ defines the longitudinal momentum fraction of the photon carried by the quark. The incident electron or projectile nucleus propagates along the $+z$ direction with momentum $P^+ p^\mu$, defined in light-cone coordinates as $p^+ \equiv (1,0, \bm{0}_\perp)/\sqrt{2}$. 

The photon flux $f_\gamma(x_\gamma)$ parameterizes the quasi-real photon distribution. For the UPC scenario, the photon flux generated by the projectile nucleus must be integrated over the transverse impact parameter space, avoiding hadronic overlap, $[2R_A, \infty)$, where $R_A$ denotes the nuclear radius. Derived via classical electrodynamics, the resulting collinear photon distribution reads~\cite{Bertulani:1987tz, Bertulani:2005ru, Baltz:2007kq}
\begin{eqnarray}
x_\gamma f_\gamma(x_\gamma)= \frac{2 Z^2 \alpha_{e}}{\pi} \left [  \zeta K_0(\zeta) K_1(\zeta) -\frac{\zeta^2}{2} \left ( K_1^2(\zeta)-K_0^2(\zeta) \right ) \right ],
\end{eqnarray}
with $\zeta \equiv 2x_\gamma M_p R_A$, $M_p$ representing the proton mass, and $Z$ the atomic number. The functions $K_0(\zeta)$ and $K_1(\zeta)$ are modified Bessel functions of the second kind. Kinematically, the photon momentum fraction is reconstructed from the rapidities and the jet transverse momentum as $x_\gamma =\frac{P_\perp}{\sqrt{s}}(e^{y_1}+e^{y_2})$, where $\sqrt{s}$ is the center-of-mass energy per nucleon pair. Conversely, for $ep$ collisions, the leading-order QED photon PDF within the electron is formulated as
\begin{eqnarray}
   f_\gamma(x_\gamma,\mu^2)=   \frac{\alpha_e}{2\pi}  \frac{1+(1-x_\gamma)^2}{x_\gamma} \ln \frac{ \mu^2}{x_\gamma^2 m_e^2}. 
  \end{eqnarray}
Here, $m_e$ is the electron mass and $\mu$ represents the factorization scale. While computed at leading order, this formulation guarantees sufficient precision for the present phenomenological scope. For higher-precision studies, DGLAP evolution can be implemented to evaluate the photon PDF at arbitrary factorization scales (see Ref.~\cite{Liu:2020rvc} for a detailed discussion).

The diffractive gluon TMD of the pomeron, $G_{\mathbb{P}}(x_g, x_{\mathbb{P}}, q_\perp)$, admits a clear probability interpretation. It describes the likelihood of finding a gluon with momentum fraction $x=x_g/x_{\mathbb{P}}$ relative to the pomeron, which itself carries a momentum fraction $x_{\mathbb{P}}$ of the parent nucleon. These variables are constrained by the external kinematics as $x_g = x_{q\bar{q}} = P_\perp(e^{-y_1} + e^{-y_2})/\sqrt{s}$ and $x_{\mathbb{P}} = x_{q\bar{q}} + l_\perp e^{-y_3}/\sqrt{s}$.  In the CGC formalism, the pomeron gluon distribution is related to the dipole-nucleus scattering amplitude via the unintegrated gluon distribution (UGD)~\cite{Iancu:2021rup, Iancu:2022lcw, Iancu:2023lel}:
\begin{eqnarray}\nonumber
 x_gG_{\mathbb{P}}(x_g,x_{\mathbb{P}} ,q_\perp)\!   =\! \frac{S_\perp (N_c^2-1)}{8\pi^4 (1\! -x) }  \left [  \frac{x q_\perp^2}{1\!-\!x} \int \! r_\perp d r_\perp J_2(q_\perp r_\perp) K_2 \left (\! \sqrt{\frac{x q_\perp^2 r_\perp^2  }{1-x}} \right ) {\cal T}_g(x_{\mathbb{P}}, r_\perp) \right]^2, \\\label{pgluonTMD}
\end{eqnarray}
where $S_\perp$ is the transverse area of the nucleus. The dipole amplitude ${\cal T}_g(x_{\mathbb{P}}, r_\perp)$ encodes the non-linear QCD dynamics and can be computed using the McLerran-Venugopalan (MV) model or parametrized via the Golec-Biernat-Wüsthoff (GBW) model. In the dilute limit (large $q_\perp$), this distribution exhibits a characteristic $1/q_\perp^4$ power-law scaling, indicating that the typical transverse momentum of the third semi-hard jet naturally aligns with the saturation scale $Q_s$.

\subsection{TEEC}

We now reformulate the acoplanarity observable within the TEEC framework~\cite{Gao:2019ojf}. By measuring the transverse momentum weighted angular correlation between particle pairs, the TEEC provides a theoretically controlled alternative to conventional jet definitions. Specifically, the transverse-momentum weighting of final-state particles makes the TEEC infrared and collinear safe and reduces the dependence on jet clustering and axis definitions.

For diffractive dijet production, the TEEC is uniquely suited to probe the azimuthal decorrelation between the two leading jets. The kinematic mapping between the TEEC scaling variable $\tau$ and the acoplanarity $\phi_\perp$ is established via a geometric expansion. In the nearly back-to-back limit, this mapping simplifies to
\begin{eqnarray}
  \tau=\frac{1+\cos (\phi_1-\phi_2)}{2}  \approx \frac{\phi_\perp^2}{4}=\frac{q_x^2}{4P_\perp^2}.
\end{eqnarray}
Exploiting this relation, the TEEC distribution can be written in the factorised form 
\begin{align}
  \frac{d\Sigma}{d \tau } \!
  = \! \int \!\! d^2 P_\perp dy_1 dy_2  \frac{P_\perp}{\sqrt{\tau} } \ \sigma_0  x_\gamma f_\gamma(x_\gamma)  \int  &
  \frac{d b_x}{2\pi} e^{2i b_x \sqrt{\tau} P_\perp } e^{-  \mathrm{Sud}_a(b_x)}  x_g G_{\mathbb{P}}(x_g,x_{\mathbb{P}},b_x)\notag \\
  & \times J_q^{\rm NP}(b_x)J_{\bar q}^{\rm NP}(b_x).
  \label{eq:teec}
\end{align}
This formula isolates the partonic hard scattering $\sigma_0$, the soft-gluon resummation, and the non-perturbative structural distributions. Crucially, the diffractive gluon distribution in impact parameter space, which encapsulates the non-linear saturation dynamics of the target, is obtained via a one-dimensional Fourier transform of the UGD as
\begin{eqnarray}
  x_g G_{\mathbb{P}}(x_g,x_{\mathbb{P}},b_x) =\int d q_x' dq_y' \ 
  e^{-i b_x  q_x'}  x_g G_{\mathbb{P}}(x_g,x_{\mathbb{P}},q_\perp').
\end{eqnarray}
The all-order resummation of soft-gluon emissions within this formulation is governed by both perturbative and non-perturbative Sudakov elements. The NLL contribution to the perturbative Sudakov exponent, $\mathrm{Sud}_a(b_x)$, is evaluated as~\cite{Gao:2019ojf, Gao:2023ivm}
\begin{equation}\label{eq:suda_DL}
  \mathrm{Sud}_a(b_x)=\int_{\mu_{bx}}^{P_\perp} \frac{d \mu }{\mu} \frac{\alpha_s }{\pi} \left [ C_A \ln\left(\frac{P_\perp^2}{\mu^2}\right)-2C_A \beta_0 +2C_F \ln\left(\frac{P_\perp^2}{\mu^2}\right)-3C_F \right ],
\end{equation}
where $\beta_0=11/12-N_f/18$ and the integration is truncated at the conjugate infrared scale $\mu_{bx}=2 e^{-\gamma_E}/|b_x|$. 

To rigorously model the transition into the non-perturbative domain at large $b_x$, we introduce a TEEC-specific non-perturbative Sudakov form factor, $J_q(b_x)$. Parameterized from global phenomenological fits~\cite{Kang:2023oqj}, it reads
\begin{equation}
  J_q^{\rm NP}(b_x,Q_0)=\exp \left [ -S_{\text{NP}}^{\text{TEEC}}(b_x)\right ]=\exp \left [ -g_0^e \sqrt{|b_x|} -g_1^e |b_x| -g_2^e b_x^2\right ],
\end{equation}
where the extracted coefficients are $g_0^e=0.226~\text{GeV}^{1/2}$, $g_1^e=0.463~\text{GeV}$, and $g_2^e=0.033~\text{GeV}^2$. This factor successfully absorbs the intrinsic transverse momentum fluctuations of final fragmenting partons.

\subsection{Dijet decorrelation}

It is important to emphasize that, in the context of dijet production, the introduction of TEEC is not strictly necessitated by infrared safety. Unlike di-hadron correlations, which are sensitive to fragmentation functions, the standard dijet azimuthal decorrelation is itself a robust, infrared-safe observable that is largely insensitive to hadronization effects. In this work, we discuss both observables on an equal footing to exploit their complementary strengths. The primary motivation for analyzing the standard dijet decorrelation lies in its sensitivity to the jet axis definition. By contrasting the SJA with the WTA scheme, we can probe their distinct responses to soft gluon radiation. This comparison provides a valuable handle on non-perturbative physics and serves as a rigorous test of the resummation formalism, particularly in distinguishing soft radiation within the jet cone from global recoil effects.

Consistent with the 1D factorization theorem, the differential cross section mapped to the azimuthal decorrelation angle $\phi_{\perp}=q_x/P_\perp$ is analytically expressed as,
\begin{eqnarray}
  \frac{d\sigma}{d \phi_\perp }
  = \int d^2 P_\perp dy_1 dy_2\,P_\perp \, \sigma_0  x_\gamma f_\gamma(x_\gamma)   \int 
  \frac{d b_x}{2\pi} e^{i b_x q_x } e^{-  \mathrm{Sud}_a(b_x)}  x_g G_{\mathbb{P}}(x_g,x_{\mathbb{P}},b_x).
  \label{eq:phiperp}
  \end{eqnarray}
The shape of the azimuthal decorrelation depends strongly on the choice of jet axis, as different algorithms respond differently to soft recoil. We investigate two standard choices: the WTA axis and the SJA. In the WTA scheme~\cite{Bertolini:2013iqa, Neill:2016vbi}, the jet axis aligns with the most energetic constituent. This makes the axis direction insensitive to soft radiation inside the jet cone. As a result, the corresponding Sudakov factor receives only the recoil contribution that changes the inter-jet momentum imbalance, while soft radiation inside the measured jet cone does not shift the jet direction~\cite{Chien:2020hzh, Chien:2022wiq, Fu:2024fgj, Fang:2023thw, Fang:2024auf, Fu:2026nkd}
\begin{align}\label{eq:suda_DL_wta}
{\rm Sud}^{\rm WTA}_a(b_x)=\int_{\mu_{bx}}^{P_\perp} \frac{d \mu }{\mu} \frac{\alpha_s }{\pi} \left [  C_A \ln\left(\frac{P_\perp^2}{\mu^2}\right)-2C_A \beta_0 +2C_F \ln\left(\frac{M^2}{\mu^2}\right)-3C_F \right ],
\end{align}
where $M$ denotes the invariant mass of the dijet system. In contrast, the SJA, typically defined using the E-scheme in the standard anti-$k_T$ algorithm, is determined by the vector sum of the momenta of all constituents within the jet radius $R$. It is therefore sensitive to recoil from soft emissions inside the jet cone. As a result, the SJA Sudakov factor is obtained from the WTA result by subtracting the contribution of final state radiation resolved within the jet cone~\cite{Dasgupta:2001sh, Gao:2023ulg}
\begin{align}\label{eq:suda_DL_sja}
{\rm Sud}^{\rm SJA}_a(b_x)={\rm Sud}^{\rm WTA}_a(b_x)-&\left\{\int_{\mu_{bx}}^{P_\perp R} \frac{d \mu }{\mu} \frac{\alpha_s }{\pi} \left [  2C_F \ln\left(\frac{P_\perp^2R^2}{\mu^2}\right)-3C_F \right ] \right. \notag \\
&\left.+ 2 C_F C_A \frac{\pi^2}{3} u^2 \frac{1+(a u)^2}{1+(b u)^c}\right\} \theta(P_\perp R-\mu_{bx}).
\end{align}
with $u=\ln[\alpha_s(\mu_{bx})/\alpha_s(P_\perp R)]/(11-2/3n_f)$, $a=0.85 \, C_A$, $b=0.86\,C_A$ and $c=1.33$ \cite{Dasgupta:2001sh}. The Heaviside step function ensures that this subtraction is strictly restricted to emissions bounded by the jet radius $R$.

\subsection{Heavy quark pair decorrelation}

The diffractive production of heavy-quark pairs ($c\bar{c}$ or $b\bar{b}$) offers a powerful probe of the saturation regime, serving as a natural complement to the light dijet channel. The heavy-quark mass $m_Q$, $m_Q\ll P_\perp$, introduces an additional semi-hard scale that guarantees the reliability of perturbative QCD even at moderate transverse momenta, effectively insulating the observable from non-perturbative soft interactions. This makes heavy-quark pair production a comparatively clean channel for probing DTMDs. The finite quark mass fundamentally modifies the contributions of soft-gluon radiation. Driven by the dead-cone effect, collinear gluon emissions are dynamically suppressed at forward angles $\theta \lesssim m_Q/E$, where $E$ denotes the energy of the radiating heavy quark. Within the Sudakov resummation framework, this physical suppression dictates that the collinear logarithmic evolution must be explicitly truncated at the mass scale $m_Q$. Analytically, the heavy-quark Sudakov form factor is obtained from the massless WTA baseline by subtracting the collinear radiative phase space bounded by $\mu < m_Q$ \cite{vonKuk:2023jfd, vonKuk:2024uxe, Marquet:2025jdr, Dai:2026hso, Gao:2026azd} 
\begin{align}\label{eq:suda_DL_HF}
{\rm Sud}^{\rm HQ}_a(b_x)={\rm Sud}^{\rm WTA}_a(b_x)-\int_{\mu_{bx}}^{m_Q} \frac{d \mu }{\mu} \frac{\alpha_s }{\pi} \left [  2C_F \ln\left(\frac{m_Q^2}{\mu^2}\right)-C_F \right ] \theta(m_Q-\mu_{bx}).
\end{align}
This subtraction systematically diminishes the overall magnitude of the Sudakov exponent. We note that the derivation of Sudakov form factors for massive particles has been extensively validated in analogous processes, notably in dilepton production via photon–photon fusion in UPCs~\cite{Klein:2018fmp, Klein:2020jom, Hatta:2021jcd, Shao:2023zge, Shao:2023bga, Shi:2024gex}.

\section{Phenomenological results}\label{sec:num}

With the theoretical framework and resummation formalism established, we now proceed to the numerical evaluation of the azimuthal observables. The primary objective is to quantify the effects of soft-gluon radiation and to determine the sensitivity of these observables to saturation dynamics across varying kinematic regimes. In this section, we present numerical predictions for the light-quark TEEC, the heavy-quark pair acoplanarity ($\phi_\perp$), and the dijet decorrelations under the SJA and WTA definitions.

We conduct a comprehensive analysis across three distinct experimental environments: ultra-peripheral $\text{Pb-Pb}$ collisions at the LHC, $e\text{Pb}$ collisions at the future EIC, and $ep$ collisions at HERA. This complementary set of colliding systems allows us to systematically probe saturation physics and evaluate the nuclear modification factor over a broad range of $x$ kinematics and target species.

Our numerical implementation incorporates both perturbative and non-perturbative components into the Sudakov form factor. To regulate the non-perturbative effects in the large impact parameter region ($|b_x| \sim 1/\Lambda_{\text{QCD}}$), we adopt the standard $b_*$-prescription, defining $b_* = |b_x| / \sqrt{1 + b_x^2/b_{\text{max}}^2}$ with $b_{\text{max}} = 1.5~\text{GeV}^{-1}$. The total Sudakov factor is thus modified to
\begin{align}
    \text{Sud}(b_x) = \text{Sud}_{a}(\mu_{b_*}) + S_{\text{NP}}(b_x),
\end{align}
where the non-perturbative input $S_{\text{NP}}(b_x)$ is taken from fits to SIDIS and Drell-Yan data~\cite{Su:2014wpa,Echevarria:2020hpy}:
\begin{equation}
S_{\text{NP}}(b_x) = g_1 b_x^2 + g_2 \ln\left(\frac{Q}{Q_0}\right) \ln\left(\frac{|b_x|}{b_*}\right).
\end{equation}
Here, the fitted parameters are $g_1 = 0.106~\text{GeV}^2$, $g_2 = 0.84$, and $Q_0^2 = 2.4~\text{GeV}^2$, while $Q$ represents the relevant hard scale of the process (identified here as $P_\perp$).

The diffractive scattering off the saturated nucleus is governed by the dipole amplitude ${\cal T}_g(x_{\mathbb{P}}, r_\perp)$. To model this interaction, we employ the GBW parametrization~\cite{Golec-Biernat:1998zce} as
\begin{eqnarray}
 {\cal T}_g(x_{\mathbb{P}}, r_\perp) = 1 - \exp \left [  -\frac{1}{4} Q_s^2(x_{\mathbb{P}}) r_\perp^2  \right].
\end{eqnarray}
Within this model, the proton saturation scale is parameterized as 
\begin{equation}
Q_{sp}^2(x) = (x_0/x)^\lambda Q_{s0}^2,
\end{equation}
with $Q_{s0}^2=1 \, \text{GeV}^2$, $x_0 = 3.04 \times 10^{-4}$, and $\lambda = 0.288$. For a heavy nucleus of atomic number $A$, the saturation momentum scales as $Q_{sA}^2 = c A^{1/3} Q_{sp}^2$, where the coefficient $c \sim 0.5$ accounts for the effective centrality dependence.

\begin{figure}[t]
    \centering
    \includegraphics[width=0.48\linewidth]{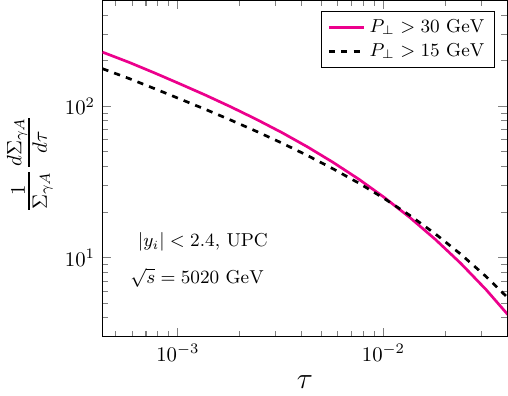}
    \includegraphics[width=0.48\linewidth]{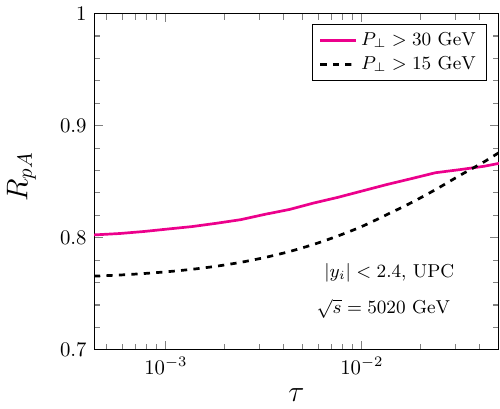}
    \caption{Left: The normalized TEEC distribution in $PbPb$ UPCs at $\sqrt{s} = 5.02$ TeV. Right: The nuclear modification factor $R_{pA}$ for the TEEC distribution.}
    \label{fig:UPC_res}
\end{figure}

\begin{figure}[t]
    \centering
    \includegraphics[width=0.48\linewidth]{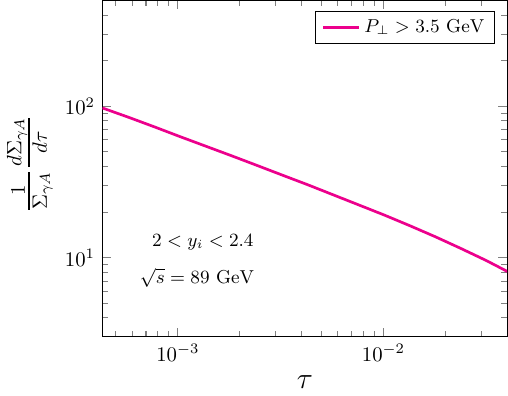}
    \includegraphics[width=0.48\linewidth]{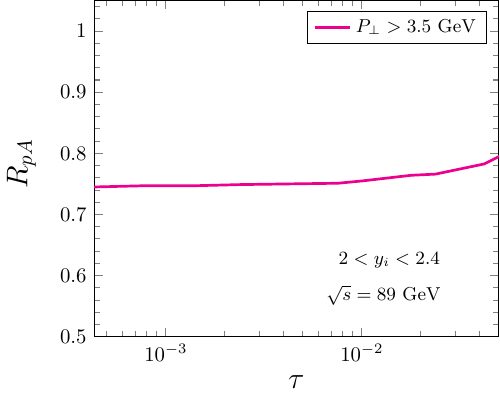}
    \caption{Left: The normalized TEEC distribution for $eA$ collisions at the EIC ($\sqrt{s} = 89$ GeV). Right: The nuclear modification factor $R_{pA}$.}
    \label{fig:EIC_res}
\end{figure}

\begin{figure}[t]
    \centering
    \includegraphics[width=0.5\linewidth]{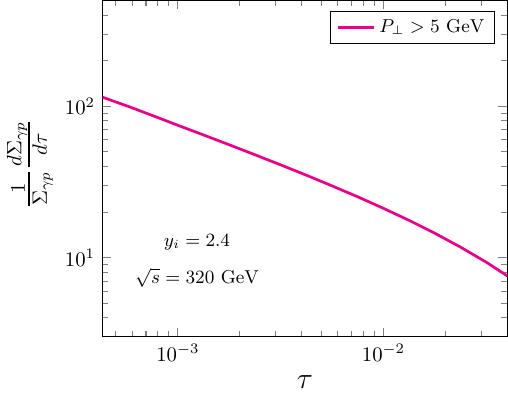}
    \caption{The normalized TEEC distribution for diffractive dijet photoproduction at HERA ($\sqrt{s} = 320$ GeV).}
    \label{fig:HERA_res}
\end{figure}

We begin our phenomenological study with diffractive dijet production in $\text{Pb-Pb}$ UPCs at $\sqrt{s} = 5.02~\text{TeV}$. We restrict the jet rapidities to $|y_{1,2}| < 2.4$ and test two distinct kinematic cuts: $P_\perp > 15~\text{GeV}$ and $P_\perp > 30~\text{GeV}$. The left panel of Fig.~\ref{fig:UPC_res} displays the predicted TEEC distributions. Both kinematic selections yield a pronounced peak in the back-to-back limit ($\tau \rightarrow 0$). Notably, the distribution for the $30~\text{GeV}$ threshold is slightly narrower than that for $15~\text{GeV}$. Although a larger hard scale inherently amplifies soft-gluon radiation and the corresponding Sudakov broadening, this effect is overpowered by the geometric $1/P_\perp$ suppression intrinsic to the angular definition $\phi_\perp \approx q_\perp / P_\perp$. Therefore, the higher-$P_\perp$ dijet event leads to a more sharply peaked TEEC angular distribution.

The right panel of Fig.~\ref{fig:UPC_res} illustrates the corresponding  nuclear modification factor, which is defined as
\begin{equation}
 R_{pA}(\tau) =\frac{1}{A} \frac{\frac{ d\sigma^{\gamma A}}{d\tau} }{\frac{ d\sigma^{\gamma p}}{d\tau} }. 
\end{equation}
The ratio exhibits a clear suppression at small $\tau$ and an enhancement in the tail region at larger $\tau$. This redistribution of the TEEC, from back-to-back peak into the broader tail, is a defining hallmark of saturation physics. It is due to the enhanced transverse momentum broadening induced by the larger nuclear saturation scale ($Q_{sA} > Q_{sp}$).

We subsequently extend our predictions to the future EIC at $\sqrt{s} = 89~\text{GeV}$, selecting events with $P_\perp > 3.5~\text{GeV}$ and photon virtualities characteristic of quasi-real photoproduction (see Fig.~\ref{fig:EIC_res}). While the qualitative features of the TEEC distribution and its associated nuclear modification factor closely mirror the LHC results, the lower center-of-mass energy and reduced $P_\perp$ threshold probe a distinctly different region of the pomeron momentum fraction $x_{\mathbb{P}}$. This kinematic shift provides complementary constraints on the energy evolution of non-linear saturation dynamics. 

The EIC will play an indispensable role by providing a pristine experimental environment for diffractive deep inelastic scattering, entirely free from the factorization-breaking complications inherent to hadronic collisions. The predicted $R_{pA}$ demonstrates a robust sensitivity to the nuclear saturation scale, firmly establishing the TEEC at the EIC as a highly discriminatory probe of non-linear QCD dynamics. Furthermore, as illustrated in Fig.~\ref{fig:HERA_res}, analogous TEEC distributions evaluated for $ep$ collisions at the HERA energy of $\sqrt{s} = 320~\text{GeV}$ provide a vital perturbative baseline. This baseline serves as a critical consistency check for the theoretical formalism before extending its application to the complex dynamics of complex nuclear targets.

\begin{figure}[t]
    \centering
    \includegraphics[width=0.48\linewidth]{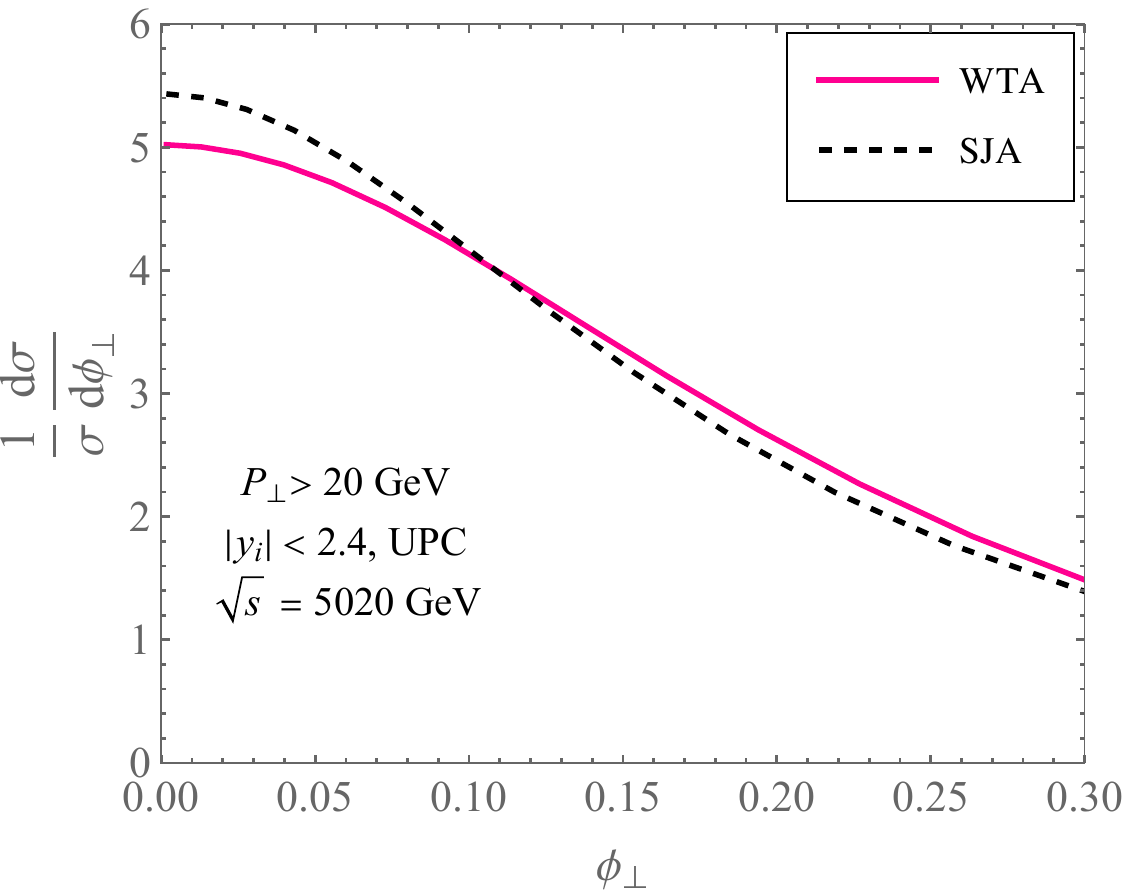}
    \includegraphics[width=0.48\linewidth]{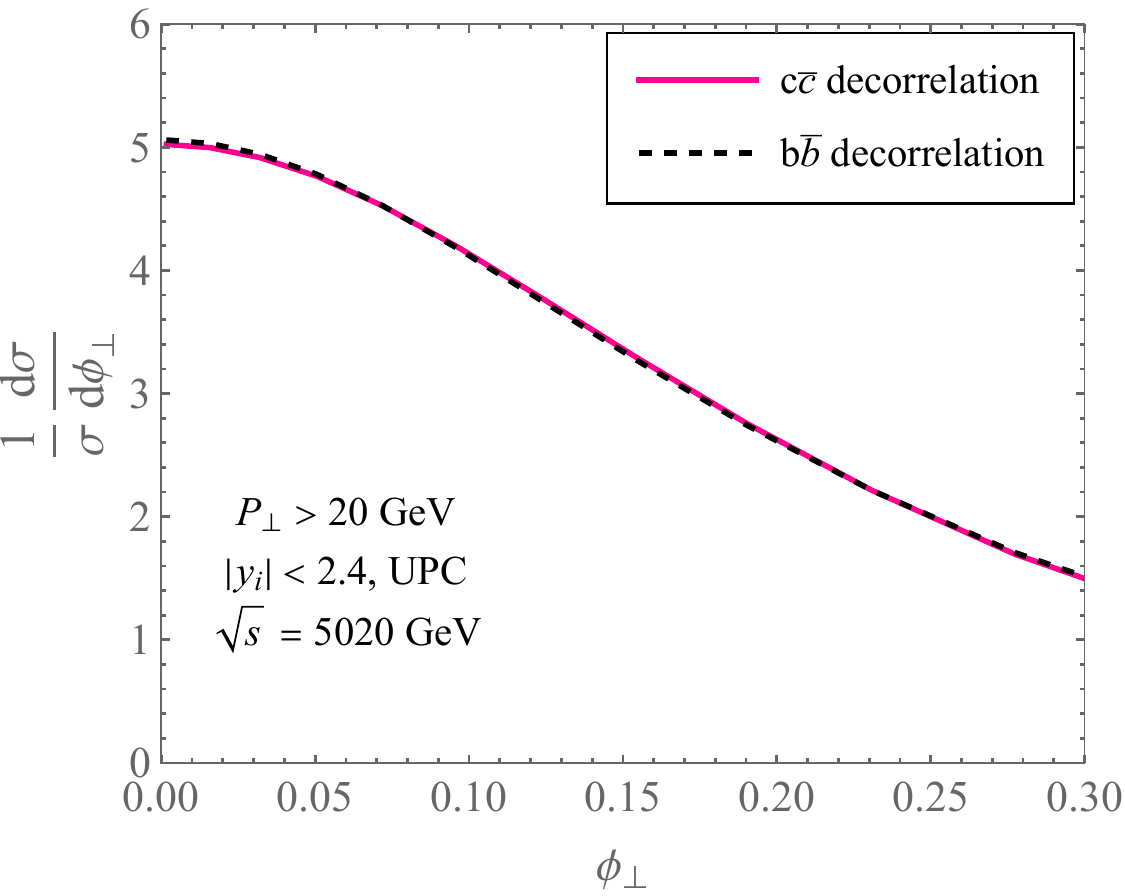}\\[2mm]
    
    \includegraphics[width=0.48\linewidth]{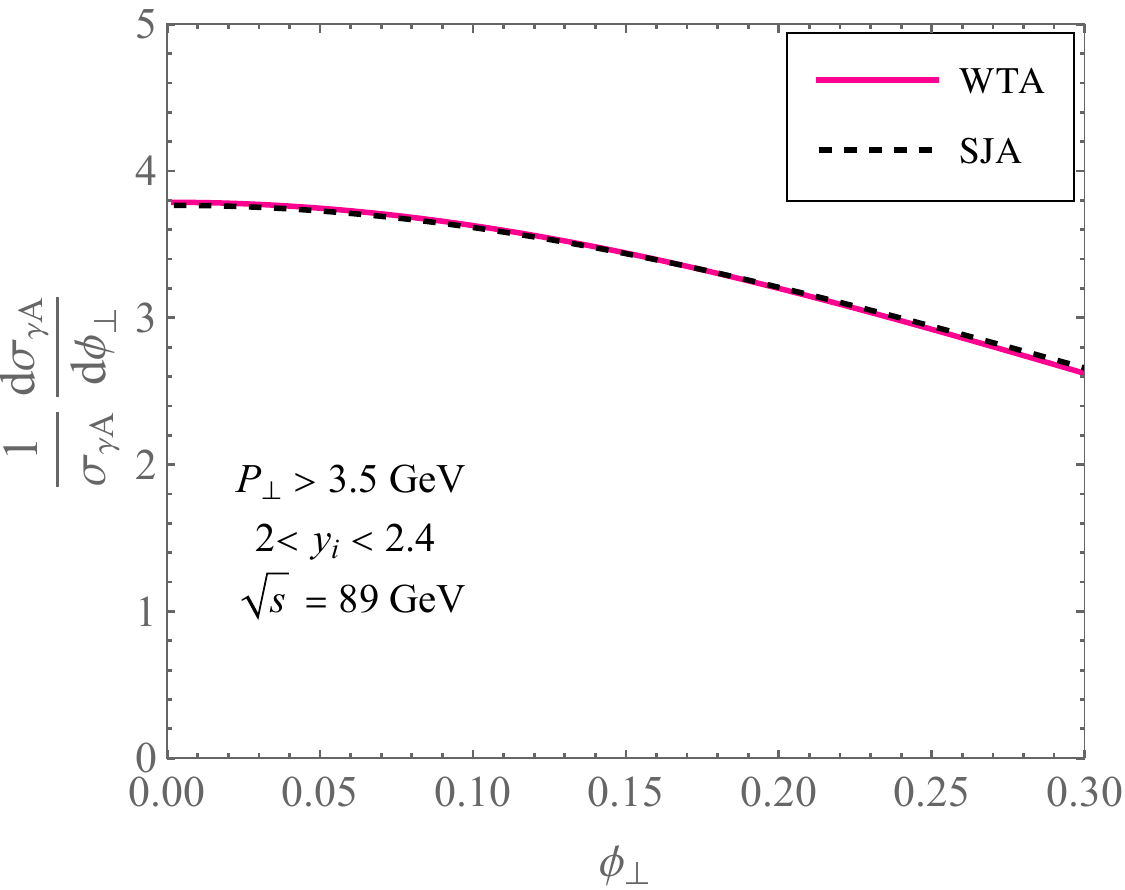}
    \includegraphics[width=0.48\linewidth]{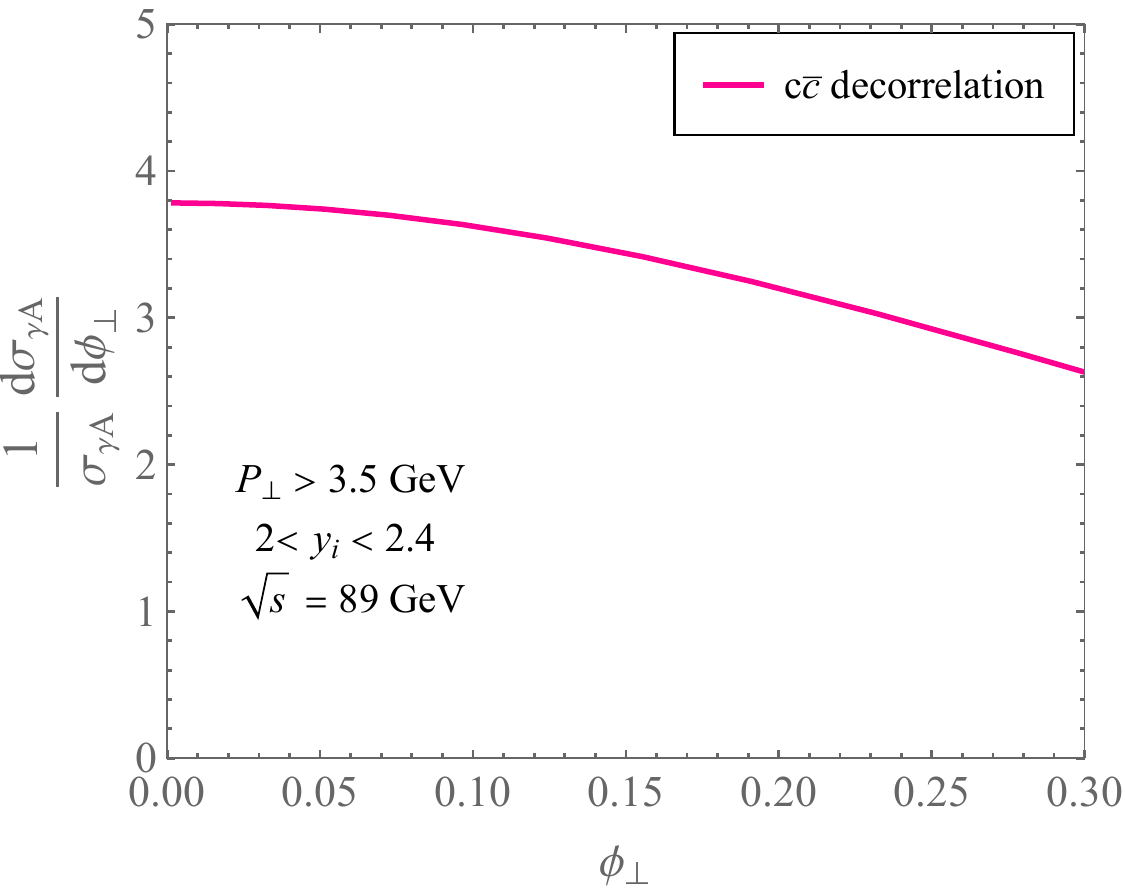}\\[2mm]
    
    \includegraphics[width=0.48\linewidth]{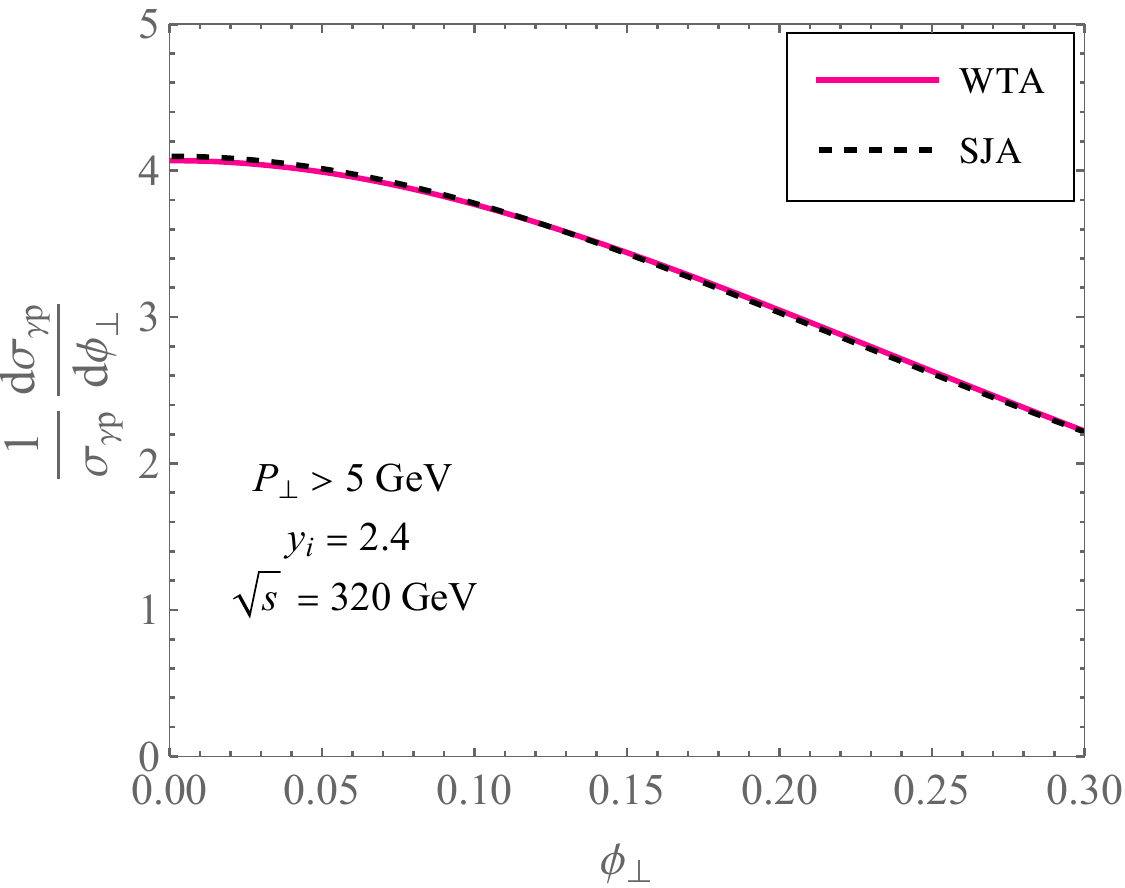}
    \includegraphics[width=0.48\linewidth]{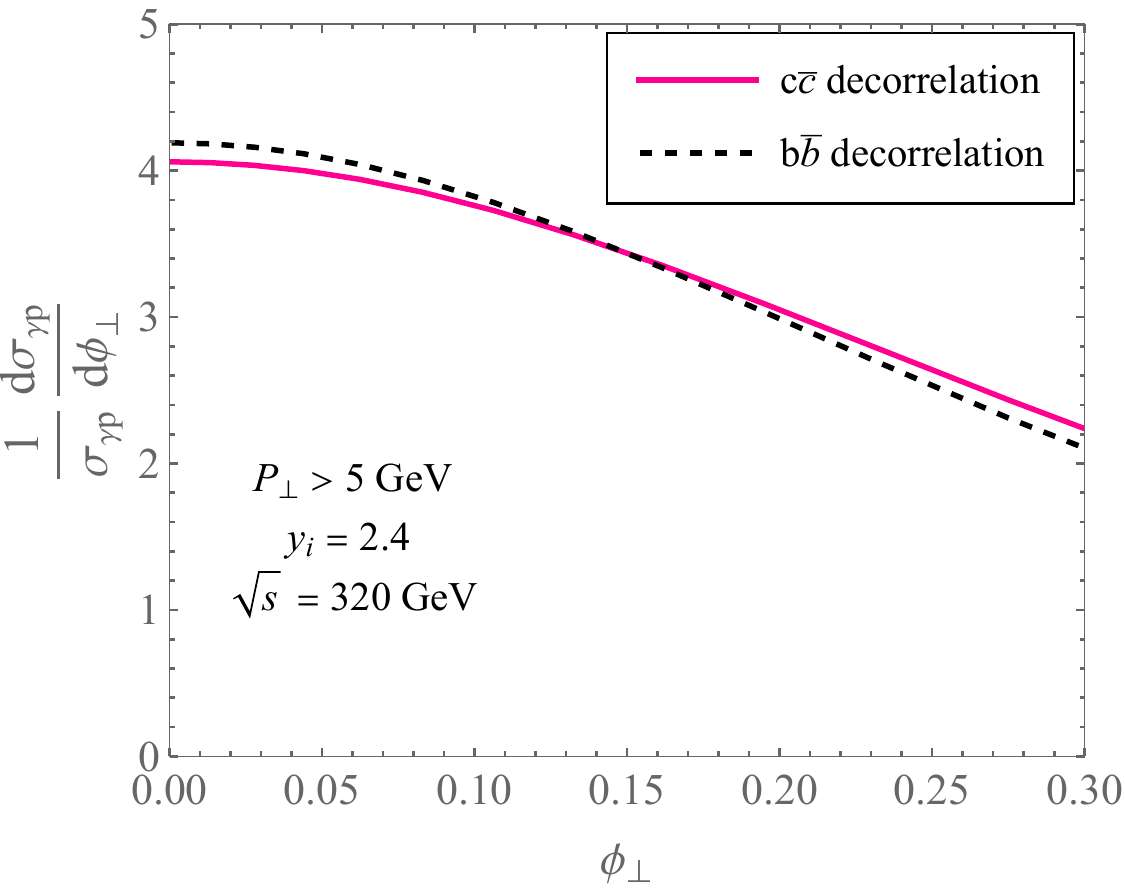}
    
    \caption{The dijet decorrelation distribution comparing the WTA and SJA jet algorithms (left column), and the corresponding heavy-quark pair decorrelation (right column) across different collider environments. Top row: $\text{Pb-Pb}$ UPCs at $\sqrt{s} = 5.02~\text{TeV}$. Middle row: $e\text{A}$ collisions at the EIC energy of $\sqrt{s} = 89~\text{GeV}$. Bottom row: $ep$ collisions at the HERA energy of $\sqrt{s} = 320~\text{GeV}$. The jet radius R is taken to be 0.6.}
    \label{fig:dijet_HQ_all}
\end{figure}

In addition to the TEEC, investigating standard dijet observables and heavy-flavor production provides valuable complementary measurements to probe the underlying saturation dynamics. Figure~\ref{fig:dijet_HQ_all} shows the azimuthal decorrelation distributions ($\phi_\perp$) across three distinct collider environments: Pb-Pb UPCs at $\sqrt{s} = 5.02~\text{TeV}$, $eA$ collisions at the EIC with $\sqrt{s} = 89~\text{GeV}$, and $ep$ collisions at HERA with $\sqrt{s} = 320~\text{GeV}$. The left column compares the light dijet decorrelation reconstructed within the WTA and SJA schemes. 

Complementing the light-jet analysis, the right column presents the azimuthal decorrelation for heavy-quark pairs ($c\bar{c}$ and $b\bar{b}$). Driven by the dead-cone effect, collinear gluon radiation is dynamically suppressed at angles smaller than $m_Q/E$. In our factorization framework, this physical suppression truncates the collinear logarithmic evolution in the Sudakov factor at the heavy quark mass scale $m_Q$. Consequently, the restricted phase space for collinear radiation leads to a much narrower decorrelation for heavy-quark pairs compared to light dijets. Furthermore, the heavier $b\bar{b}$ pairs show a slightly narrower distribution than the $c\bar{c}$ pairs due to their larger mass. These consistent features across diverse collision systems confirm the robustness of the DTMD resummation formalism. More importantly, they establish heavy-quark pair production as a clean probe to isolate the genuine saturation signal from perturbative radiative effects.

\section{Conclusion}
\label{sec:conclusion}

In this paper, we have studied azimuthal angular correlations in diffractive dijet production, a process suitable for probing gluon DTMDs. While diffractive exclusive dijet production has long been considered a golden channel for accessing the gluon Wigner distribution, recent theoretical advances have highlighted the dominance of the semi-inclusive (tri-jet) channel in UPCs and photoproduction processes. This channel, characterized by the emission of an additional semi-hard gluon, evades color transparency suppression and introduces new dynamical effects. We have presented an analysis of the TEEC for light quarks, the $\phi_\perp$ azimuthal correlations for heavy-quark pairs, and dijet observables defined using the SJA and WTA within the framework of diffractive TMD factorization. By performing an all-order resummation of soft gluon logarithms, we have provided predictions for the TEEC distribution in UPCs at the LHC, as well as for HERA and future EIC. A central finding of our work is that ISR, uniquely present in the semi-inclusive channel due to the color-octet nature of the $q\bar{q}$ pair, acts as a significant source of azimuthal decorrelation. This effect broadens the azimuthal distribution.

To further disentangle the effect of soft gluon radiation, we extended our theoretical framework to investigate the impact of different jet axis definitions and heavy quark mass effects. Furthermore, our analysis of heavy quark pairs ($c\bar{c}$ and  $b\bar{b}$) revealed that the dead-cone effect suppresses collinear emissions, thereby truncating the Sudakov evolution and resulting in a narrower azimuthal decorrelation compared to light di-hadrons. This mass-dependent scaling offers a highly perturbative and complementary probe of the nuclear saturation scale.

Furthermore, our predictions highlight the indispensable role of the EIC. Unlike hadronic collisions, where factorization breaking complicates the interpretation, the EIC provides a clean laboratory for these studies. We find that the TEEC observable at the EIC is sensitive to the nuclear saturation scale, with the nuclear modification factor $R_{pA}$ serving as a robust probe of non-linear QCD dynamics in the heavy nucleus.

In summary, the acoplanarity of diffractive dijet production may serve as a relatively clean and experimentally practical observable for diffractive physics. We have taken advantage of the fact that the two-particle acoplanarity $\phi_\perp$ appears to be a more suitable kinematic variable than the momentum imbalance $q_\perp$, as it makes use of the high angular resolution of detectors to help mitigate the systematic limitations associated with limited jet energy resolution. This choice could allow for a measurement of the Pomeron’s intrinsic momentum structure with less smearing. By establishing a perturbative baseline that includes the important effects of initial state radiation, jet kinematics, and heavy quark dead-cone effects, this work hopes to contribute to future efforts to extract the gluon Wigner distribution and to study saturation physics at the high-energy frontier. Future work might focus on incorporating higher-order corrections to the hard and soft functions, which would help to further reduce theoretical uncertainties.

\vspace{3mm}
\noindent{\it Acknowledgments.}
The authors thank Shu-Yi Wei for helpful discussions. This work has been supported by the National Natural Science Foundations of China under Grant No.~12175118 (J.Z.), No.~12321005 (J.Z.), No.~12575091 (J.Z.), No.~12275052 (D.Y.S.), No.~12147101 (D.Y.S.), No.~12547102 (D.Y.S.), No.~12405151 (C.Z.) and No.~12475084 (Y.Z.). D.Y.S. is also supported by the Innovation Program for Quantum Science and Technology under grant No. 2024ZD0300101. Y.Z. is also supported by the Natural Science Foundation of Shandong Province (Grant No. ZR2024MA012).

\bibliographystyle{JHEP}
\bibliography{ref.bib}

\end{document}